\documentclass[pre,twocolumn,showpacs,preprintnumbers,amsmath,amssymb,floatfix]{revtex4}

\usepackage{graphicx}
\usepackage{dcolumn}
\usepackage{bm}

\usepackage{verbatim,vmargin,amsmath,calc, mathrsfs}
\usepackage{amsthm, amssymb, dsfont}
\usepackage{epsfig, pstricks, pst-plot,pst-node}
\usepackage{graphicx}
\usepackage{psfrag, wasysym}

\newpsobject{showgrid}{psgrid}{subgriddiv=5}

\newcommand{\av}[1]{\langle #1 \rangle}

\newcommand{\mom}[2]{\langle #1^{#2} \rangle}

\newcommand{\oforder}[1]{\mathscr{O}(#1)}

\begin{document}

\title {Mean-Field Theory and Sandpile Models}

\author {Matthew Stapleton}
\email{ms599@imperial.ac.uk}
\affiliation{Blackett Laboratory, Imperial College London,
Prince Consort Road, London SW7 2BW, United Kingdom}
\author {Kim Christensen}
\email{k.christensen@imperial.ac.uk}
\affiliation{Blackett Laboratory, Imperial College London,
Prince Consort Road, London SW7 2BW, United Kingdom}

\date{\today}

\begin{abstract}
We review and refine the concept of a mean-field theory for the study of sandpile models,
which are of central importance in the study of self-organized criticality.
By considering the simple one-dimensional random walker with an absorbing and reflecting boundary we are able to construct
a complete mean-field picture which we can solve in detail for different types of driving.
Using this theory, we are able to clarify the effect of finite driving
rate on sandpile models, as well as the observed sensitivity of certain universal quantities on the driving.
\end{abstract}

\maketitle

Sandpile models form the core of the study of self-organized criticality (SOC)\cite{Bak,Dhar,Manna,Oslo}.
These models describe the movement of particles in bursts called avalanches, with SOC
characterized by the fact that the avalanche size probability exhibits simple finite-size scaling,
which can be described by the critical exponents $\tau$ and $\Delta$ (defined below).
These exponents have been determined exactly for very few models \cite{Dhar}, with
the exceptions generally being mean-field theories,
where it is well established that $\tau = 3/2$ and $\Delta=2$ \cite{Christensen1993,Zapperi1995}.
However, one question which has not been addressed in the literature is whether mean-field theory
can tell us anything more qualitative about the nature of sandpile models.
In the following we address this question by formulating and solving a
mean-field theory based on escape times of a random walker.
We find, of course, that there is scaling and universality in this
model, and that it arises as a direct consequence of symmetry and bulk
conservation.
We also see clearly the roles played by the exponents $\tau$ and
$\Delta$ and clarify the sensitivity of these
exponents on the driving, as observed numerically in Refs.~\cite{Nakanishi,Bengrine,Oslo, Anders, Christensen2004,Christensen2004b}.
Finally, we identify the effect of non-zero driving rates, which has been of recent interest because
of connections to absorbing state phase transitions \cite{Vespignani, Christensen2004b}, and use this
as a basis for writing down the scaling theory of SOC.

The difficulty with solving sandpile models analytically is that there is a lattice which can contain particles
which are stable as well as unstable.
During the course of an avalanche, only unstable particles move but stable particles may be made unstable
by interacting with those which are already unstable.
Thus, the stable particles form a `medium' which interacts with the avalanches formed by unstable particles.
Keeping track of these interactions soon becomes analytically intractable and thus there is a real
deficiency in exact solutions for sandpile models.
A mean-field theory replaces the medium seen by each degree of freedom with a self-consistent average,
such that fluctuations are ignored.
In this case, we look at a moving particle as it enters a site.
There will be some probability that it is made stable, as well as a probability that it excites a stable particle
already there to become unstable.
If the sandpile is in a stationary state and particles are conserved in the bulk, then the average number
of unstable particles that leave a (bulk) site is equal to the average that entered.
Hence, in the mean-field theory, we neglect the interactions of the particles entirely and unstable
particles simply perform a random walk until they leave the system through a boundary site.

To define our mean-field theory, we consider a random walker on a line, $x \in [0,L]$ with an absorbing
boundary at $x = 0$ and a reflecting boundary at $x = L$.
We introduce the walker at some point $0 < x_0 < L$ and define the `avalanche' size 
as the time, $T$, it takes for the walker to leave the system through the absorbing  boundary at $x=0$.
The probability density for the walker at time $t$, which starts at $x_0$ at $t=0$ is
$\phi \equiv \phi(x,t;x_0)$ and the dynamics are described by the diffusion equation
\begin{equation}
\frac{\partial \phi}{\partial t} = D \frac{\partial^2 \phi}{\partial x^2} - v \frac{\partial \phi}{\partial x},\label{eq:diffusion}
\end{equation}
where $D$ is the diffusion constant and $v$ is the drift velocity.
The probability density for $T$ is simply given by the current out of the system at time $t=T$:
\begin{equation}
P(T;L) \equiv D \left.\frac{\partial \phi(x,T;x_0)}{\partial x}\right|_{x=0}.\label{eq:derivative}
\end{equation}
As well as being a mean-field theory for sandpile models like those in Refs.~\cite{Bak,Dhar,Manna,Oslo}, this model can be justified
as a mean-field theory of the Eulerian walker \cite{Eulerian}
and is in the same universality class as the critical branching process where each
tree is restricted to a maximum of $L$ generations \cite{Harris,Feller}.

The SOC hypothesis predicts that under certain circumstances the event-size probability
exhibits simple finite-size scaling, that is, for $T \gg 1$ and $L \gg 1$,
\begin{equation}
P(T;L) = a T^{-\tau} \mathscr{G}(b T/L^{\Delta}),\label{eq:MFFSS}
\end{equation}
where $a$ and $b$ are non-universal constants, $\tau$ and $\Delta$ are universal exponents
and $\mathscr{G}$ is a scaling function.
We can also study the moments of this probability density
\begin{equation}
\mom{T}{n} = \int^{\infty}_0\ T^nP(T;L)dT
\end{equation}
which will scale with system size
\begin{equation}
\mom{T}{n} \approx \Gamma_n L^{\gamma_n} \quad \text{for } L \gg 1,
\end{equation}
where $\Gamma_n$ are non-universal amplitudes and $\gamma_n = \Delta(1+n-\tau)$.
For the mean-field theory we will identify the reason for this scaling,
which may give us clues to the cause of scaling in SOC systems.
We shall also yield some explanation for why in SOC
systems we find that $\tau$ depends on the location of the driving, $x_0$, whereas $\Delta$
remains constant:
Stationarity gives us the scaling relations (for instance, see Ref.~\cite{Nakanishi})
\begin{equation}
\Delta(2 - \tau) = \begin{cases}1 \qquad x_0 = \text{const.}\\
2 \qquad x_0 / L = \text{const.}
\end{cases}
\end{equation}
and it is found numerically that $\Delta$ is the same for both cases whereas $\tau$ changes
\cite{Nakanishi,Bengrine,Oslo, Anders, Christensen2004,Christensen2004b}.
This is important to understand as we might expect from the renormalization group that universal quantities such as critical exponents
are only affected by symmetry-breaking or non-conservative perturbations.

For now we set the drift velocity $v=0$, anticipating the fact that scaling is observed only in this case.
We shall handle the finite drift case later in this letter.
If the diffusers start close to the absorbing boundary then they take a time $\approx L^2/D$ to reach the reflecting 
boundary on the right hand side and for small times the system does not `feel' that the system is finite.
Hence, when $t \ll L^2/D$, $\phi(x,t;x_0)$ takes the same form as the propagator for diffusion on the semi-infinite line,
\begin{equation}
\phi \approx \frac{1}{\sqrt{4\pi Dt}}\left[ e^{-\frac{(x-x_0)^2}{4Dt}} - e^{-\frac{(x+x_0)^2}{4Dt}} \right].
\end{equation}
For $t \gg L^2/D$, only the lowest order mode of $\phi(x,t;x_0)$ survives and
\begin{equation}
\phi \approx \frac{\pi^2}{2L^3} x_0 x e^{-\frac{\pi^2}{4L^2}Dt}.
\end{equation}
Taking the derivative in Eq.~\eqref{eq:derivative}, we find
\begin{equation}
\!\!P(T;L) \approx \begin{cases} \frac{x_0}{\sqrt{4\pi D}} T^{-\frac{3}{2}} \quad &x_0^2/D \ll T \ll L^2/D\\
\frac{\pi^2 D}{2L^3} x_0 e^{-\frac{\pi^2}{4L^2}DT} \quad &T \gg L^2/D
\end{cases}
\end{equation}
which can be put in the scaling form
\begin{equation}
P(T;L) \approx x_0 T^{-\frac{3}{2}} \mathscr{G}(DT/L^2)\label{eq:scalingdist}
\end{equation}
with $\mathscr{G}(x) = \text{const.}$ for small arguments and $x^{3/2}e^{-x}$ for
large arguments.
For constant $x_0$, this is simple finite-size scaling and by comparing to Eq.~\eqref{eq:MFFSS} we identify
$\Delta = 2$ and $\tau = 3/2$.
In the case of bulk driving, $x_0 = y_0 L$, where $y_0$ is a constant,
we have to recast the equation
\begin{align}
P(T;L) &\approx y_0 L T^{-3/2} \mathscr{G}(DT/L^2)\nonumber \\
&\equiv y_0T^{-1} \tilde{\mathscr{G}}(DT/L^2)\label{eq:solnPscal}
\end{align}
where $\tilde{\mathscr{G}}(x) = x^{-1/2} \mathscr{G}(x)$.
Hence, for $x_0 \propto L$, we still have $\Delta = 2$ but $\tau = 1$.

We also find the scaling behavior of the moments.
This can be done by writing down the generating function, $Z \equiv Z(\lambda,x_0;L)$,
where
\begin{equation}
\mom{T}{n} = (-1)^n \left.\frac{\partial^n Z}{\partial \lambda^n}\right|_{\lambda=0}.
\end{equation}
This is easily determined to be
\begin{align}
Z = 
2\frac{L^{2n}}{D^n}
\frac{\sinh\sqrt{\lambda}\cosh\left((1-\frac{x_0}{L})\sqrt{\lambda}\right)}{\sinh2\sqrt{\lambda}}.
\end{align}
We note that $\mom{T}{n}$ has dimensions of $(\text{time})^n$ and dimensional analysis tells us that 
\begin{equation}
\mom{T}{n} = \frac{L^{2n}}{D^n} g_n(x_0/L)
\end{equation}
where $g_n$ is a scaling function for the $n$th moment of the dimensionless ratio $x_0/L \leq 1$.
If we write down $g_n$ as a power series in $x_0/L$, we have for $n > 0$,
\begin{equation}
\mom{T}{n} = \frac{L^{2n}}{D^n} \sum^{\infty}_{m=1} C_{m,n} \left(\frac{x_0}{L}\right)^m\label{eq:momentsleading}
\end{equation}
where the coefficients $C_{m,n}$ are numbers independent of $L$, $D$ and $x_0$.
It is clear that the leading order behavior in $L$ comes from the
lowest order term, so that for $n > 0$,
\begin{equation}
\mom{T}{n} = x_0 \frac{L^{2n-1}}{D^{n}} C_{1,n} + \oforder{L^{2n-2}}.\label{eq:solnscaling}
\end{equation}
We can find the $C_{1,n}$ directly from $Z$:
\begin{equation}
C_{1,n} = (-1)^{n-1} n! \frac{2^{2n}(2^{2n}-1)B_{2n}}{(2n)!},
\end{equation}
where $B_n$ are Bernoulli numbers.

From Eq.~\eqref{eq:momentsleading} we can write down the scaling of the moments in the usual
scaling form,
\begin{equation}
\mom{T}{n} = \Gamma_n L^{\gamma_n} + \oforder{L^{\gamma_n - 1}}\label{eq:sum}
\end{equation}
with $\gamma_n = \Delta(1+n-\tau)$ where $\Delta = 2$ is the gap exponent and $\tau = 3/2$ is the escape time
exponent.
The amplitudes $\Gamma_n$ are non-universal and are equal to
\begin{equation}
\Gamma_n = \frac{x_0}{D^{2n}} (-1)^{n-1} n! \frac{2^{2n}(2^{2n}-1)B_{2n}}{(2n)!}.
\end{equation}
Note that for bulk driving $x_0/L = \text{const.}$, Eq.~\eqref{eq:solnscaling} implies that $\tau$ changes its value to $1$, but
$\Delta$ remains unchanged, in agreement with Eq.~\eqref{eq:solnPscal}.

We also consider the moment ratios
\begin{equation}
g_n \equiv \frac{\mom{T}{n}\av{T}^{n-2}}{\mom{T}{2}^{n-1}} = \frac{\Gamma_n \Gamma_1^{n-2}}{\Gamma_2^{n-1}}
\end{equation}
which are defined for $n > 0$.
For an event-size probability given by Eq.~\eqref{eq:MFFSS}, the $g_n$ approach universal values for $L \gg 1$.
However, we see in Eq.~\eqref{eq:momentsleading} that for $x_0/L = \text{const.}$, 
each term in the expansion of $\mom{T}{n}$ will be of the same order in
$L$, and so the amplitude $\Gamma_n$ will be the sum of all the terms and not just the first.
We therefore expect universal amplitude ratios to differ for the two types of driving.

We can therefore see that the reason for the dependence on $\tau$ and not $\Delta$ on the driving
is because the probability density, $P(T;L)$, and therefore its moments are proportional to $x_0$ to lowest order.
In terms of the exponents $\gamma_n$, this means that the gap between the $\gamma_n$, which is proportional to $\Delta$, is unchanged.
We argue that the linearity in $x_0$ can be shown to be a consequence of conservation of particles, 
as well as a symmetry between $x$ and $x_0$ in diffusion: $\phi(x,t;x_0)$ not only obeys
the diffusion equation written in Eq.~\eqref{eq:diffusion}, but also
\begin{equation}
\frac{\partial \phi}{\partial t} = D \frac{\partial^2 \phi}{\partial x_0^2} + v \frac{\partial \phi}{\partial x_0}.\label{eq:diffusionbackward}
\end{equation}
One interpretation of this is as a time reversal symmetry:
the probability of a particle diffusing from $x$ to $y$ in time $t$ is identical to that for $y$ to $x$,
providing we reverse the direction of the drift.
Even the presence of absorbing boundary conditions does not break this symmetry.
The effect of this is clear.
Symmetry and conservation constrain the dependence of the observables on $x_0$ to be the same
as for $x$, which is linear to lowest order.

There has been much recent interest in studying the effects of finite drive rate on the behavior
of sandpile models.
It is widely held that scaling only occurs when the driving rate is set to zero, with a finite
drive rate introducing a cutoff length scale for avalanche sizes.
In our mean-field theory, we wait until each walker leaves the system before adding the next,
which corresponds to an infinitely slow driving rate.
Consider now the situation where we add walkers at some finite rate $q$
and measure the event sizes as the time between instances of having zero walkers in the system.
This situation is modeled well by treating $q$ like a drift, $q \rightarrow -v$.
Taking the solution in the limit $L \rightarrow \infty$, we find for $T \gg x_0^2/D$,
\begin{equation}
P(T;q) \approx \frac{x_0}{\sqrt{4\pi D}} T^{-\frac{3}{2}} e^{-\frac{q^2}{4D}T}.
\end{equation}
Hence, there is an exponential cutoff with the characteristic time
\begin{equation}
T_{q} = \frac{4D}{q^2}
\end{equation}
and the model exhibits scaling only when $T_q \rightarrow \infty$, that is, $q \rightarrow 0$.
This is in agreement with the idea that SOC is only
achieved in the limit of zero drive rate \cite{Vespignani}.

We are now in a position to formulate a general scaling theory for the moments of the event size
probability in sandpile models.
We have a number of length scales, $L$, $x_0$ and $D/q$, which we have already introduced, plus
an additional length scale corresponding to details of fluctuations on the lattice, $a$.
Writing $\mom{T}{n}$ in terms of dimensionless quantities, we have
\begin{equation}
\mom{T}{n} = \frac{L^{2n}}{D^n} \Phi_n\left(\frac{x_0}{L},\frac{a}{L}, \frac{q L}{D}\right)
\end{equation}
where $\Phi_n$ is some scaling function.
The scaling function may scale with $a$, but we do not expect it to produce a cutoff in the scaling
function.  Hence, we write
\begin{equation}
\mom{T}{n} = \frac{L^{2n}}{D^n} \left(\frac{a}{L}\right)^{\eta(n)} \Xi_n\left(\frac{x_0}{L}, \frac{q L}{D}\right)
\end{equation}
where $\eta(n)$ is an exponent which may depend on $n$ and $\Xi_n$ is the scaling function $\Phi_n$ with the
$a$ dependence factored out.
For sandpiles with conservation we anticipate similar constraints on $x_0$ as for the mean-field
model and hence we make the conjecture
\begin{equation}
\mom{T}{n} = x_0 L^{2n-\eta(n) - 1} \frac{a^{\eta(n)}}{D^n} \Psi_n\left(\frac{q L}{D}\right)
\end{equation}
where, again, $\Psi_n$ is obtained from $\Xi_n$.
It is now clear that when $q \rightarrow 0$, we lose another
length scale and $\mom{T}{n}$ exhibit pure scaling.

We have investigated the mean-field theory of sandpiles
in some detail in order to gain insight into the nature of the
``criticality'' of SOC sandpiles.
We have found that a simple mean-field theory exhibits scaling in the presence of bulk
conservation and slow driving.
We have shown that critical exponents and amplitude ratios,
are dependent on the type of driving used, a feature which has been
observed numerically in non-trivial sandpile models.
It should be noted that this mean-field description depends only on conservation and
symmetry in the underlying sandpile model, and does not describe any mechanism of self-organization
or fluctuations in the medium, which are the unique properties of sandpiles.
However, it is remarkable that such a simple mean-field theory has been able to capture so successfully the
remaining features of these non-trivial models.

\end{document}